\documentclass[aps,prl,reprint,superscriptaddress,longbibliography,floatfix]{revtex4-2}

\usepackage{graphicx}
\graphicspath{{images/}}

\usepackage{bm}
\usepackage{dsfont}
\usepackage[dvipsnames]{xcolor}
\usepackage{pstricks}
\usepackage{verbatim}
\usepackage{units}
\usepackage{natbib}
\usepackage{mathtools}
\usepackage{multirow}
\usepackage{enumitem}
\usepackage{mathrsfs}
\usepackage{leftidx}
\usepackage{xspace}
\usepackage{amsmath,mathtools,amsfonts,amssymb}
\usepackage[normalem]{ulem} 
\usepackage{soul}
\usepackage{physics}
\usepackage{tabularx}
\usepackage{upgreek}
\usepackage[normalem]{ulem} 
\newcolumntype{Y}{>{\centering\arraybackslash}X}

\usepackage{hyperref}
\hypersetup{colorlinks=true,linktoc=all,linkcolor=blue,breaklinks=true,citecolor=blue,urlcolor=blue}

\usepackage{braket}

\usepackage[english]{babel}
\addto\captionsenglish{}

\makeatletter
\renewcommand\paragraph{\@startsection{paragraph}{4}{\z@}%
  {3.25ex \@plus1ex \@minus.2ex}%
  {-0em}%
  {\normalfont\normalsize\itshape\indent}}
\begin{document}

\title{Thermalization and dephasing in collisional reservoirs }

\author{Jorge Tabanera-Bravo}
\email{jorgetab@ucm.es}
\affiliation{Departamento de Estructura de la Materia, F\'isica T\'ermica y Electr\'onica and GISC, Universidad Complutense de Madrid, Pl. de las Ciencias 1. 28040 Madrid, Spain}

\author{Juan M. R. Parrondo}
\email{parrondo@ucm.es}
\affiliation{Departamento de Estructura de la Materia, F\'isica T\'ermica y Electr\'onica and GISC, Universidad Complutense de Madrid, Pl. de las Ciencias 1. 28040 Madrid, Spain}

\author{Massimiliano Esposito}
\email{massimiliano.esposito@uni.lu}
\affiliation{Complex Systems and Statistical Mechanics, Physics and Materials Science Research Unit, University of Luxembourg, L-1511 Luxembourg, G.D. Luxembourg}

\author{Felipe Barra}
\email{fbarra@dfi.uchile.cl }
\affiliation{Departamento de F\'isica, Facultad de Ciencias F\'isicas y Matem\'aticas, Universidad de Chile, 837.0415 Santiago, Chile}

\date{\today}
\begin{abstract}
We introduce a wide class of quantum maps that arise in collisional reservoirs and are able to thermalize a system if they operate in conjunction with an additional dephasing mechanism. These maps describe the effect of collisions and induce transitions between populations that obey detailed balance, but also create coherences that prevent the system from thermalizing. We combine these maps with a unitary evolution acting during random Poissonian times between collisions and causing dephasing. 
We find that, at a low collision rate, the nontrivial combination of these two effects causes thermalization in the system. This scenario is suitable for modeling collisional reservoirs at equilibrium. We justify this claim by identifying the conditions for such maps to arise within a scattering theory approach and provide a thorough characterization of the resulting thermalization process.
\end{abstract}
\maketitle

\setlength{\tabcolsep}{4pt}
\renewcommand{\arraystretch}{1.3}

Collisional reservoirs are becoming an essential tool for the study of open quantum systems and quantum thermodynamics   
\cite{Strasberg2017,Guarnieri2020,Rodrigues2019,Ciccarello2022}. The term applies to situations in which a system interacts sequentially with 
the internal degrees of freedom of particles extracted from a reservoir. The effect of each interaction is described by a quantum map that can be obtained under some simplifying assumptions, like supposing that the particle and the system interact for a given time. However, this approach turns out to be thermodynamically inconsistent in certain situations of interest because switching on and off the interaction involves the performance of a work that, for example, prevents the system from thermalizing when the reservoir is a thermal bath at equilibrium \cite{Barra2015,Strasberg2017,Jacob2021,Ciccarello2022}. This drawback imposes some limitations when applying collisional models to fundamental problems in quantum thermodynamics, such as the thermalization of spatially extended systems \cite{Hofer2017} and of systems with non-commuting conserved quantities \cite{Halpern2022,Kranzl2022,Majidy2023}.

One way to restore thermodynamic consistency is to consider the particle's spatial degrees of freedom and analyze the collision as an autonomous event. In this case, there is no longer a need for an external agent to switch on and off the interaction \cite{Jacob2021}. In this context, using a scattering theory approach, two necessary conditions for thermalization have been found: {\em i)} the velocity of the particles must be distributed according to the effusion distribution at a given temperature and {\em ii)} the interaction must be time reversible \cite{Ehrich2020,Jacob2021,Tabanera2022}. Moreover, in the quantum case, the dispersion of the momentum of the incident particles must be small enough to cancel out the coherences among the eigenstates of the system's Hamiltonian \cite{Jacob2021}. For incident wave packets with a non-negligible momentum dispersion, the collision can induce coherences that prevent the system from thermalizing. 

An open question is whether the combination of these collisions with some dephasing mechanism can yield a repeated-interaction scheme that is thermodynamically consistent. This is the question that we address in this Letter.
We first analyze the problem of thermalization in a generic repeated-interaction scheme given by a quantum map whose transition probabilities obey detailed balance but, at the same time, generate coherences that drive the system out of equilibrium. Then we apply this generic analysis to quantum maps derived within scattering theory.

Consider a quantum system with Hamiltonian $H_S$ and eigenstates $H_S\ket{j}=e_{j}\ket{j}$. 
When it interacts with an auxiliary system for a given time or is bombarded by particles coming from a reservoir, the density matrix of the system changes as $\rho \to \rho'={\mathbb S}\rho$. The super-operator ${\mathbb S}$  can be written in tensorial form ${\mathbb S}^{jk}_{j'k'}$ in the eigenbasis of $H_S$:
\begin{equation}
\rho'_{j'k'}=\sum_{jk}{\mathbb S}^{jk}_{j'k'}\rho_{jk}
\end{equation}
with $\rho_{jk}\equiv \braket{j|\rho|k}$.
The term  ${\mathbb S}^{jj}_{j'j'}$ is the transition probability from eigenstate $\ket{j}$ to $\ket{j'}$. In this Letter, we analyze quantum maps with transition probabilities that obey detailed balance:
\begin{equation}\label{db}
e^{-\beta e_{j}}{\mathbb S}^{jj}_{j'j'}=
e^{-\beta e_{j'}}{\mathbb S}^{j'j'}_{jj}
\end{equation}
with respect to an inverse temperature $\beta$.

Such a map ${\mathbb S}$ can induce thermalization by itself. A relevant example is when the only nonzero entries of the tensor are the transition probabilities ${\mathbb S}^{jj}_{j'j'}$ and ${\mathbb S}^{jk}_{jk}$, with $|{\mathbb S}^{jk}_{jk}|<1$ for $j\neq k$. Then coherences (the off-diagonal entries of the density matrix $\rho$ in the eigenbasis of $H_S$) decay and populations (the diagonal entries of $\rho$ in the eigenbasis of $H_S$) evolve as a Markov chain and thermalize. This is the case of systems bombarded by wave packets with a small momentum dispersion \cite{Jacob2021,Tabanera2022}.

However, in some relevant situations, such as systems bombarded by broad packets in momentum representation coming from thermal baths at equilibrium, populations can couple to coherences through nonzero terms ${\mathbb S}^{jj}_{j'k'}$ with $j'\neq k'$. In these cases, even though the transition probabilities obey detailed balance, coherences prevent the system from thermalizing. Concatenation of the map ${\mathbb S}$ is not enough to thermalize the system and must be complemented by a dephasing mechanism. The most trivial one consists of intercalating a full dephasing superoperator ${\mathbb D}$ that kills all off-diagonal terms of the density matrix. In the eigenbasis of the Hamiltonian:
\begin{equation}\label{dephas0}
{\mathbb D}^{jk}_{j'k'}=\delta_{jj'}\delta_{kk'}\delta_{jk}.
\end{equation}
It is straightforward to prove that the detailed balance condition \eqref{db} and the complete dephasing ${\mathbb D}$ are sufficient to thermalize the system for any initial condition $\rho_0$:
\begin{equation}\label{dephas}
\lim_{n\to\infty}[{\mathbb D}{\mathbb S}]^{n}\rho_0=\frac{e^{-\beta H_S}}{Z}\equiv \rho_{\rm Therm},
\end{equation}
$Z={\rm Tr}[e^{-\beta H_S}]$ being the partition function.

However, to devise realistic scenarios, one has to consider more specific dephasing mechanisms. A candidate is the random phase added to the off-diagonal terms of the density matrix if the system is bombarded at random times and evolves under the Hamiltonian $H_S$ between collisions \cite{Ciccarello2022period}. The density matrix after $n$ collisions is
\begin{equation}\label{concat_tau}
\rho_{n}\equiv {\mathbb S}{\mathbb U}^{\tau_n}{\mathbb S}{\mathbb U}^{\tau_{n-1}}\dots {\mathbb S}{\mathbb U}^{\tau_2}{\mathbb S}{\mathbb U}^{\tau_1}\rho_{0}
\end{equation}
where ${\mathbb U}^{\tau}\rho=e^{-iH_S\tau/\hbar}\rho e^{iH_S\tau/\hbar}$ is the super-operator corresponding to the Hamiltonian unitary evolution and $\tau_1,\dots,\tau_n$ are random variables.
The density matrix $\rho(t)$ at time $t$ is given by the average
\begin{equation}
    \rho(t)=\langle \rho_n\rangle
\end{equation}
taken over all possible values of $n$ and $\tau_k$ ($k=1,2,\dots,n$) such that $t=\sum_k \tau_k$. If the collisions are Poissonian events occurring at a rate $\Gamma$, then the probability of a collision in an interval $[t,t+\Delta t]$ is $\Gamma\Delta t$, independently of past events. Hence,
\begin{equation}
    \rho(t+\Delta t)\simeq [1-\Gamma\Delta t]\,{\mathbb U}^{\Delta t}\rho(t)+\Gamma\Delta t \,{\mathbb S}\rho(t)
\end{equation}
yielding the master equation \cite{Strasberg2017,Ciccarello2022period}
\begin{equation}
    \frac{d\rho(t)}{dt}=-\frac{i}{\hbar}[H_S,\rho(t)]+\Gamma ({\mathbb S}-{\mathbb I})\rho(t).
\end{equation}
The corresponding steady state verifies
\begin{equation}
    -\frac{i}{\hbar}[H_S,\rho_{\rm ss}]+\Gamma ({\mathbb S}-{\mathbb I})\rho_{\rm ss}=0.
    \label{steady}
\end{equation}
Detailed balance for the transition probabilities \eqref{db} is not sufficient for thermalization, i.e., for having $\rho_{\rm ss}=\rho_{\rm Therm}$, even with Poissonian collisions. The reason is that, if ${\mathbb S}$ generates coherences that subsequently affect populations, then ${\mathbb S}\rho_{\rm Therm}\neq
 \rho_{\rm Therm}$. However, the generation of coherences can be reduced if the collision rate $\Gamma$ is very small. To see this, let us solve Eq.~\eqref{steady} perturbatively by inserting
\begin{equation}
    \rho_{\rm ss}=\rho^{(0)}+\Gamma \rho^{(1)}+\Gamma^2\rho^{(2)}+\dots
\end{equation}
The first-order terms yield
\begin{equation}
    \begin{split}
    -\frac{i}{\hbar}[H_S,\rho^{(0)}]&=0\\
    -\frac{i}{\hbar}[H_S,\rho^{(1)}]+({\mathbb S}-{\mathbb I})\rho^{(0)}&=0
    \\
    -\frac{i}{\hbar}[H_S,\rho^{(2)}]+({\mathbb S}-{\mathbb I})\rho^{(1)}&=0.
\end{split}
\end{equation}
Multiplying these equations by $\bra{j}$ on the left and $\ket{k}$ on the right, we get :
\begin{equation}\label{pert}
    \begin{split}
 -\frac{i}{\hbar}\Delta_{jk}\rho_{jk}^{(0)}&=0\\
    -\frac{i}{\hbar}\Delta_{jk}\rho_{jk}^{(1)}+\sum_{j'k'}{\mathbb S}_{jk}^{j'k'}\rho^{(0)}_{j'k'}-\rho_{jk}^{(0)}&=0
    \\
    -\frac{i}{\hbar}\Delta_{jk}\rho_{jk}^{(2)}+\sum_{j'k'}{\mathbb S}_{jk}^{j'k'}\rho^{(1)}_{j'k'}-\rho_{jk}^{(1)}&=0
\end{split}
\end{equation}
where $\Delta_{jk}\equiv e_j-e_k$.
For simplicity, we assume that the eigenstates of $H_S$ are non-degenerate: $\Delta_{jk}=0 \Leftrightarrow j=k$. In this case, the first equation in \eqref{pert} implies that $\rho^{(0)}$ is diagonal in the eigenbasis of $H_S$ and the second one, particularized for $k=j$, determines the diagonal terms or populations. They fulfill the following equation:
\begin{equation}
    \sum_{j'}{\mathbb S}_{jj}^{j'j'}\rho^{(0)}_{j'j'}=\rho_{jj}^{(0)}
\end{equation}
whose solution is $\rho_{\rm Therm}$ if the transition probabilities verify the detailed balance condition \eqref{db}. The off-diagonal terms are of order $\Gamma$ with
\begin{equation}
   \rho_{jk}^{(1)}=-\frac{i\hbar}{\Delta_{jk}}\sum_{j'}{\mathbb S}_{jk}^{j'j'}\rho^{(0)}_{j'j'}
   \label{eq:coherence}
\end{equation}
for $j\neq k$. The third equation in \eqref{pert} for $k=j$ determines the first-order correction to the diagonal terms:
\begin{equation}
    \sum_{j'k'}{\mathbb S}_{jj}^{j'k'}\rho^{(1)}_{j'k'}=\rho_{jj}^{(1)}
\end{equation}
which are no longer thermal. We see that the corrections to the thermal state are of order $\Gamma \hbar/|\Delta_{jk}|$. That is, the system thermalizes if the average time between collisions $1/\Gamma$ is much longer than the evolution time of the phases of the off-diagonal terms of the density matrix, which are $\hbar/|\Delta_{jk}|$.


Now we investigate the conditions under which this type of map results from the interaction between a system and a particle or unit $U$ extracted from a reservoir.

As in \cite{Jacob2021},  we study the case of a system colliding with a one-dimensional quantum particle of mass $m$. The Hamiltonian of the global setup reads
\begin{equation}\label{ham}
H=\frac{\hat p^{2}}{2m}+H_{S}+Vf(\hat x)
\end{equation}
where $\hat p$ and $\hat x$ are the momentum and position operators of the particle, respectively,   $V$ is an operator acting on the system, and $f(x)$ is a function with finite support, which is the scattering region where the system and the particle interact.
Here, we assume for simplicity that the scatterer is symmetric under spatial inversion, $f(x)=f(-x)$, and that the particle has no internal degrees of freedom. They can be incorporated in a straightforward manner, following \cite{Tabanera2022}.

We bombard the system with units prepared in a generic mix state $\rho_U$. In this case, the resulting scattering map is \cite{Taylor1972,Jacob2021}
\begin{align}
\mathbb{S}_{j'k'}^{jk} &=\sum_{\alpha=\pm}\int_{ p_{\rm inf}}^\infty dp\, \rho_{U}(p,\pi(p))\sqrt{\frac{p}{\pi(p)}}\,s_{j'j}^{(\alpha )}\left(E_p+e_j\right)\nonumber \\ &\times 
\left[s_{k'k}^{(\alpha )}(E_{p}-
\Delta_{j'j}+e_{k'})\right]^* 
\label{SY-coh}
\end{align}
where $\rho_U(p,p')=\braket{p|\rho_U|p'}$ is the density matrix of the particle in the momentum representation, $E_p=p^2/(2m)$,
$\pi(p)=\sqrt{p^{2}-2m(\Delta_{j'j}-\Delta_{k'k})}$,
and the lower limit of the integral obeys $p^2_{\rm 
inf}/(2m)=\max\{ 0,\Delta_{j'j},\Delta_{j'j}-
\Delta_{k'k}\}$. 
The quantities $s_{j'j}^{(\alpha)}(E)$ 
are the complex entries of the scattering matrix, which are related to the amplitudes of the reflecting ($\alpha=-$) and 
transmitted ($\alpha=+$) plane waves of scattering states 
\cite{Jacob2021} with total energy $E$. They are obtained by solving the 
time-independent Schr\"odinger equation for 
scattering states, which behave as plane waves 
asymptotically \cite{Jacob2021,Taylor1972} (see also \cite{Tabanera2022} for exact and approximate expressions of the scattering matrix in terms of transfer matrices).

Following similar steps as in \cite{Jacob2021,Tabanera2022}, we can obtain sufficient conditions for the scattering map \eqref{SY-coh} to obey the detailed balance condition \eqref{db}.
The first condition requires that the diagonal of the state of the particles in momentum representation coincides with the effusion distribution:
\begin{equation}\label{eff}
    \rho_{U}(p,p)=\mu_{\rm eff}(p)=\frac{\beta p}{m}e^{-\beta p^2/(2m)}.
\end{equation}
The second one is micro-reversibility:
\begin{equation}
   \label{microrev}
         s_{j'j}^{(\alpha )}(E)=s_{jj'}^{(\alpha )}(E).
\end{equation}
The proof is straightforward. Inserting \eqref{eff}  in \eqref{SY-coh} for $k=j$ and $k'=j'$, we obtain
\begin{equation}
    \mathbb{S}_{j'j'}^{jj} =\sum_{\alpha=\pm}\int_{ p_{\rm inf}}^\infty dp\, \frac{\beta p}{m}\,e^{-\beta p^2/(2m)}\,\left|s_{j'j}^{(\alpha )}\left(\frac{p^2}{2m}+e_j\right)\right|^2 
\end{equation}
with $p_{\rm inf}=\sqrt{2m\Delta_{j'j}}$ if $\Delta_{j'j}>0$ and zero otherwise. With the change of variable $\varepsilon=p^2/(2m)+e_j$, we get
\begin{equation}
    \mathbb{S}_{j'j'}^{jj} =\beta e^{\beta e_j}\sum_{\alpha=\pm}\int_{\max\{e_j,e_{j'}\}}^\infty d\varepsilon\, e^{-\beta\varepsilon} \left|s_{j'j}^{(\alpha )}(\varepsilon)\right|^2 
\end{equation}
which, together with Eq.~\eqref{microrev}, immediately yields the detailed balance condition
  \eqref{db}.

Micro-reversibility is fulfilled by the exact scattering matrix in any collision described by a Hamiltonian of the form \eqref{ham}. In \cite{Tabanera2022}, we have developed several approximations of the scattering matrix that still satisfy this condition and can be used to design simple quantum repeated-interaction thermostats.  
We use one of these approximations to analyze an explicit example below \cite{supp}.

However, if $\rho_U(p,p')\neq 0$ for $p\neq p'$, then the scattering map can create coherences \cite{Jacob2021}. For example, if $\rho_U(p,\sqrt{p^2-2m\Delta_{j'k'}})\neq 0$ for $j'\neq k'$ and the amplitudes of the transitions $j\to j'$ and $j\to k'$ are nonzero, then the term ${\mathbb S}^{jj}_{j'k'}$ is nonzero and couples the off-diagonal term $\rho_{j'k'}$ to the population $\rho_{jj}$.

In particular, for particles in a pure state, the density matrix is $\rho_U(p,p')=\phi(p)\phi^*(p')$,
where $\phi(p)$ is the wave function of the pure state in the momentum representation. Then,  a diagonal state $\rho_U$ in momentum representation can be obtained only by using plane waves with $|\phi(p)|^2\propto \delta(p-p_0)$. These plane waves are completely delocalized in space and do not induce individual collision events, but rather a continuous-time evolution of the state of the system \cite{Filippov2020}. In \cite{Jacob2021}, we have shown that this condition can be relaxed to narrow wave packets whose momentum dispersion $\sigma_p$ is small enough to avoid overlapping between outgoing packets with different energies, i.e., $\rho_U(p,\sqrt{p^2-2m\Delta_{j'k'}})\simeq 0$ for all $j'\neq k'$. On the other hand, for very broad packets, one can even obtain a unitary scattering map that preserves the entropy of the system. Consequently, in this case,  the collision is a work source from a thermodynamic point of view \cite{Jacob2022}.

We now analyze whether the dephasing induced by Poissonian collisions can restore thermalization even in the case of broad packets. As we have shown above,  this is the case if the rate of collisions is low enough. Now, we check this statement using a specific example. We consider a qubit bombarded by particles that are localized in space around an initial position $x_0$  with dispersion $\Delta x$, and whose momentum is distributed according to the effusion distribution \eqref{eff}. These two requirements can be implemented using the following Wigner function \cite{Jacob2022}:
\begin{equation}\label{wigner}
    W(p,x)=\mu_{\rm eff}(p)\,\frac{1}{\sqrt{2\pi \Delta x^2}}e^{-(x-x_0)^2/(2\Delta x^2)}
\end{equation}
which is valid if $4\pi\Delta x\sqrt{m/\beta}\geq \hbar$ \cite{Hillery1984}.
From the Wigner function, we can obtain the density matrix in momentum representation \cite{Jacob2022}:
\begin{align}
   & \rho_U(p,p') =\int dx\,W\left(\frac{p+p'}{2},x\right)e^{-i(p-p')x/\hbar}\nonumber \\
    &=\mu_{\rm eff}\left(\frac{p+p'}{2}\right)\,\exp\left[
    -\frac{\Delta x^2(p-p')^2}{2\hbar^2}-i\,\frac{(p-p')x_0}{\hbar}\right].\label{rhou}
\end{align}
Notice that the density matrix is diagonal in the momentum representation only for $\Delta x\to\infty$ (the Wigner function \eqref{wigner} in this case is not valid because the resulting state $\rho_U$ is a mixture of plane waves, which are not proper states). This is equivalent to the narrow packets considered in \cite{Jacob2021,Tabanera2022}. If one imposes the localization of the particle, which is necessary to have well-defined isolated collisions, then, the density matrix $\rho_U$ is no longer diagonal in the momentum representation. This is equivalent to the broad packets that exhibit non-negligible momentum dispersion when one localizes the packet in space, due to the Heisenberg uncertainty principle. 

The total Hamiltonian of our system is \eqref{ham}, with $H_S = (\Delta/2)\sigma_z$ and $V = \lambda\left(\sigma_x + \sigma_y \right)$, $\sigma_i$ being the three Pauli matrices of the qubit. The difference between the energy levels is $\Delta$ and $\lambda$ is the intensity of the interaction. The scattering region is the interval $[0,L]$ and $f(x)$ is the indicator function of this interval: $f(x)=1$ if $x\in [0,L]$ and zero otherwise.
\begin{figure}
    \centering
    \includegraphics[scale = 0.55]{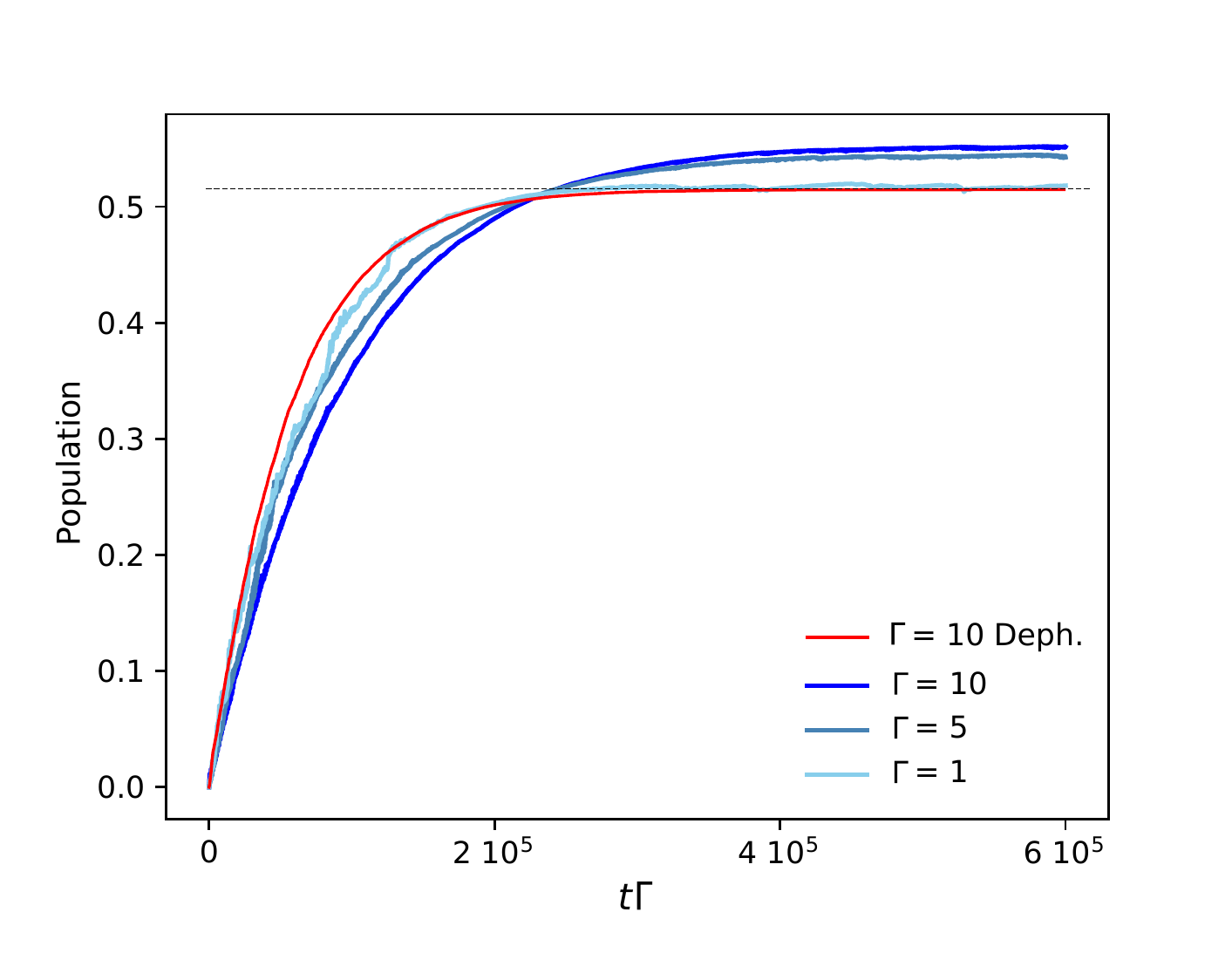}
    \caption{Time evolution of the diagonal element $\rho_{00}$ with Poissonian bombarding with rates $\Gamma = 10, 5, 1$ and when we intercalate the dephasing super-operator $\mathbb{D}$ (red). The black dashed line represents the population in the thermal equilibrium state. $\Delta = 0.6$, $\beta = 0.1$, $m = 0.1$,  $\lambda = L = \hbar = 1$, $\Delta x =1$ and $x_0 = -10$.}
    \label{fig:Dephasing}
\end{figure}

\begin{figure}[t]
    \centering
    \includegraphics[scale = .55]{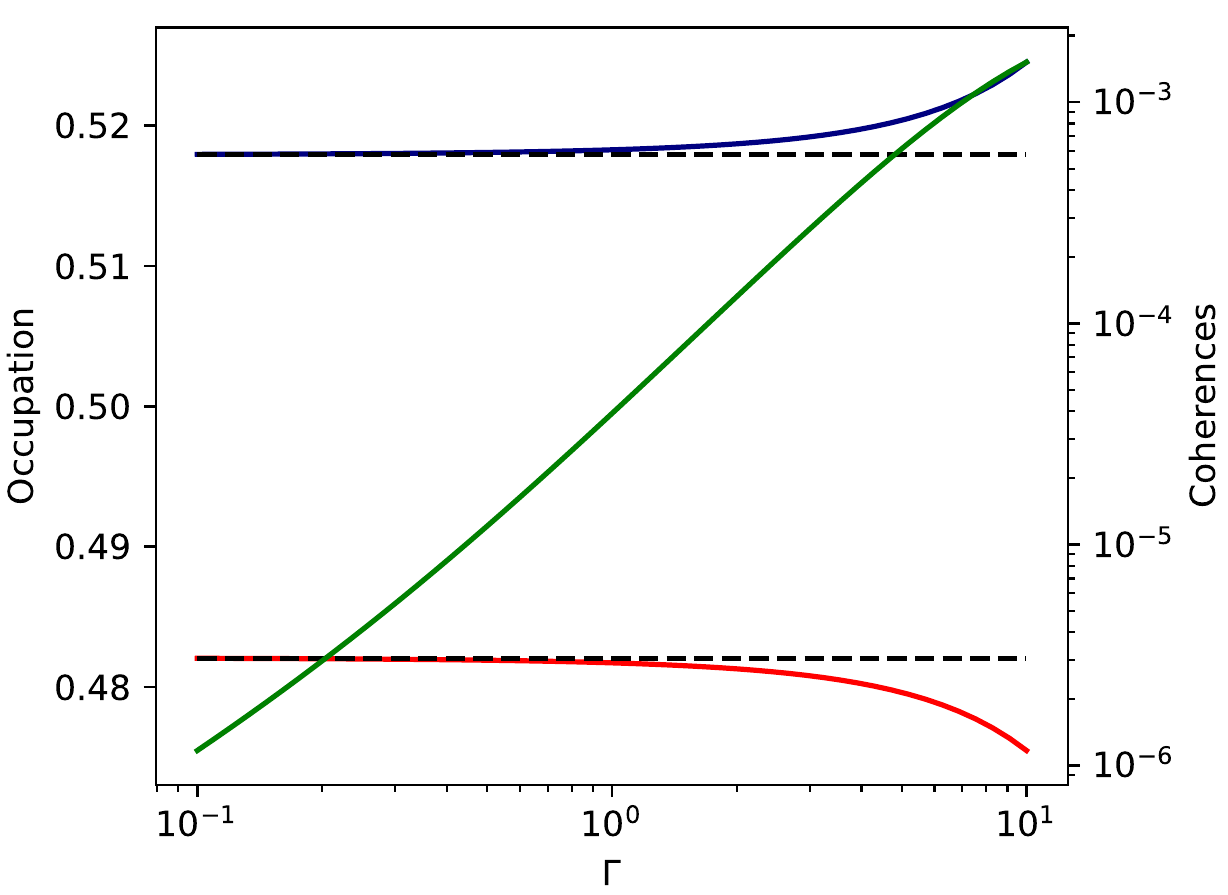}
    \caption{Numerical solution of \eqref{steady} in the case of a single qubit as a function of $\Gamma$:  $\rho_{00}$ (blue), $\rho_{11}$ (red) and $|\rho_{10}|$ (green). The black dashed lines represent the population in the thermal equilibrium state. $\Delta = 0.6$, $\beta = 0.1$, $m = 0.1$,  $\lambda = L = \hbar = 1$, $\Delta x =1$ and $x_0 = -10$.}
    \label{fig:Coherences}
\end{figure}

First, we show in Fig.~\ref{fig:Dephasing} the time evolution of the population of the ground state of the qubit when we when the evolution is given by random collisions, as in Eq.~\eqref{concat_tau} (blue curves), for different values of the bombarding rate $\Gamma$.
To obtain the time evolution, we calculate the amplitudes $s_{j'j}^{(\alpha)}(E)$ using the approximation introduced in \cite{Tabanera2022,supp}, which neglects the reflecting waves but fulfills micro-reversibility. From the scattering amplitudes and the incident density matrix \eqref{rhou}, we get the scattering map ${\mathbb S}$, using Eq.~\eqref{SY-coh}. Fig.~\ref{fig:Dephasing} shows the population of the ground state $\rho_{00}$  for a given realization of collision times $\tau_1,\tau_2,\dots$. The populations are plotted as a function of $\Gamma t$, which is the average number of collisions up to time $t$. We see that the system does not thermalize for values of $\Gamma$ well above the Bohr frequency $\Delta=0.6$ ($\hbar=1$). For comparison, we also include the population when the dephasing operator \eqref{dephas0} is applied in each collision for $\Gamma=10$. In this case, the evolution is almost deterministic and drives the qubit to the thermal state.

The steady state of the evolution can be calculated analytically by solving Eq.~\eqref{steady}. We plot the solution as a function of $\Gamma$ (for both populations and coherences) in Fig.~\ref{fig:Coherences}.
The two plots confirm our results and show that thermalization is achieved when $\Gamma$ is of the same order as the qubit frequency $\Delta/\hbar=0.6$ or lower. 


\begin{figure*}
    \includegraphics[scale = .36]{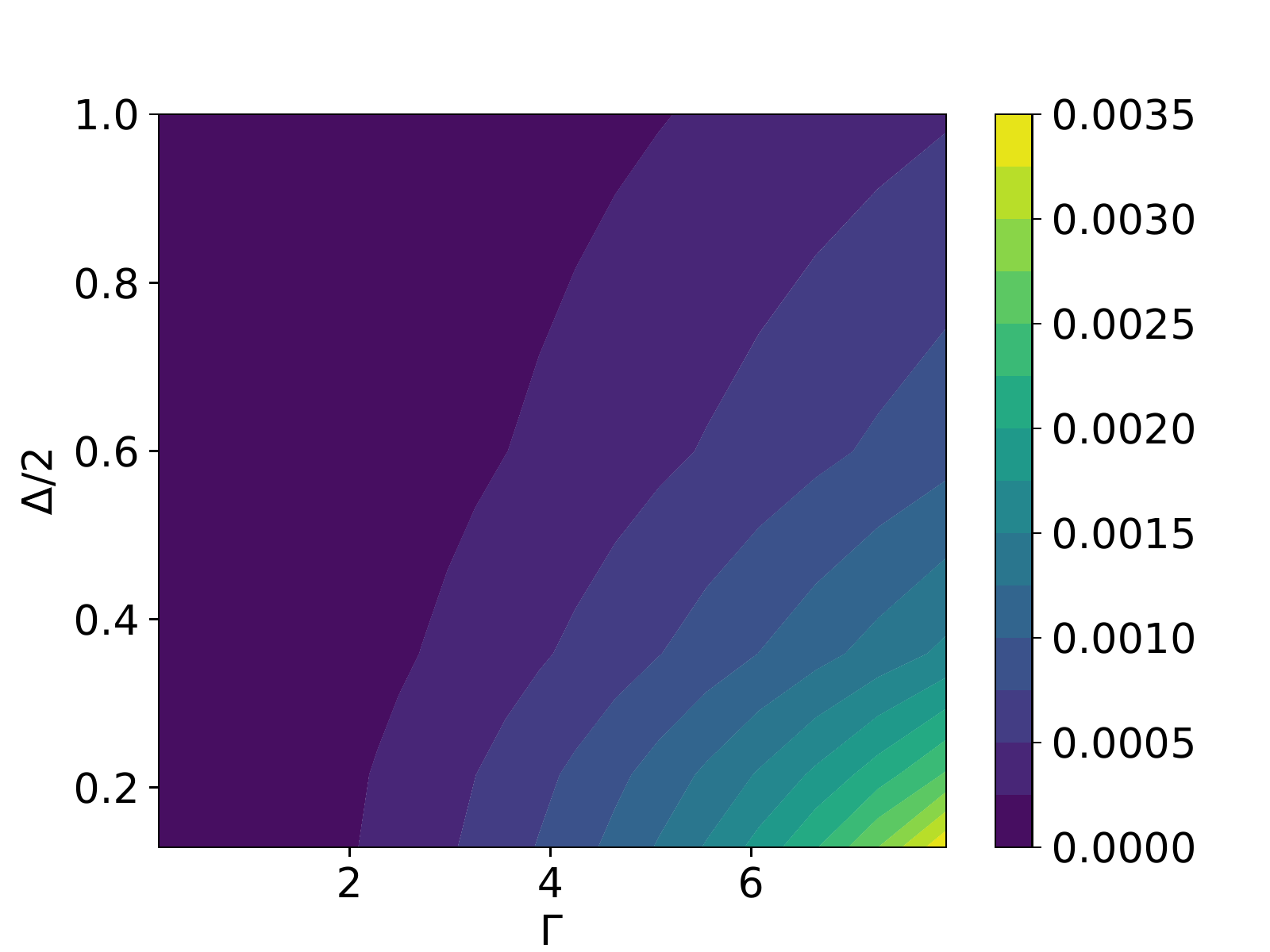}
    \includegraphics[scale = .36]{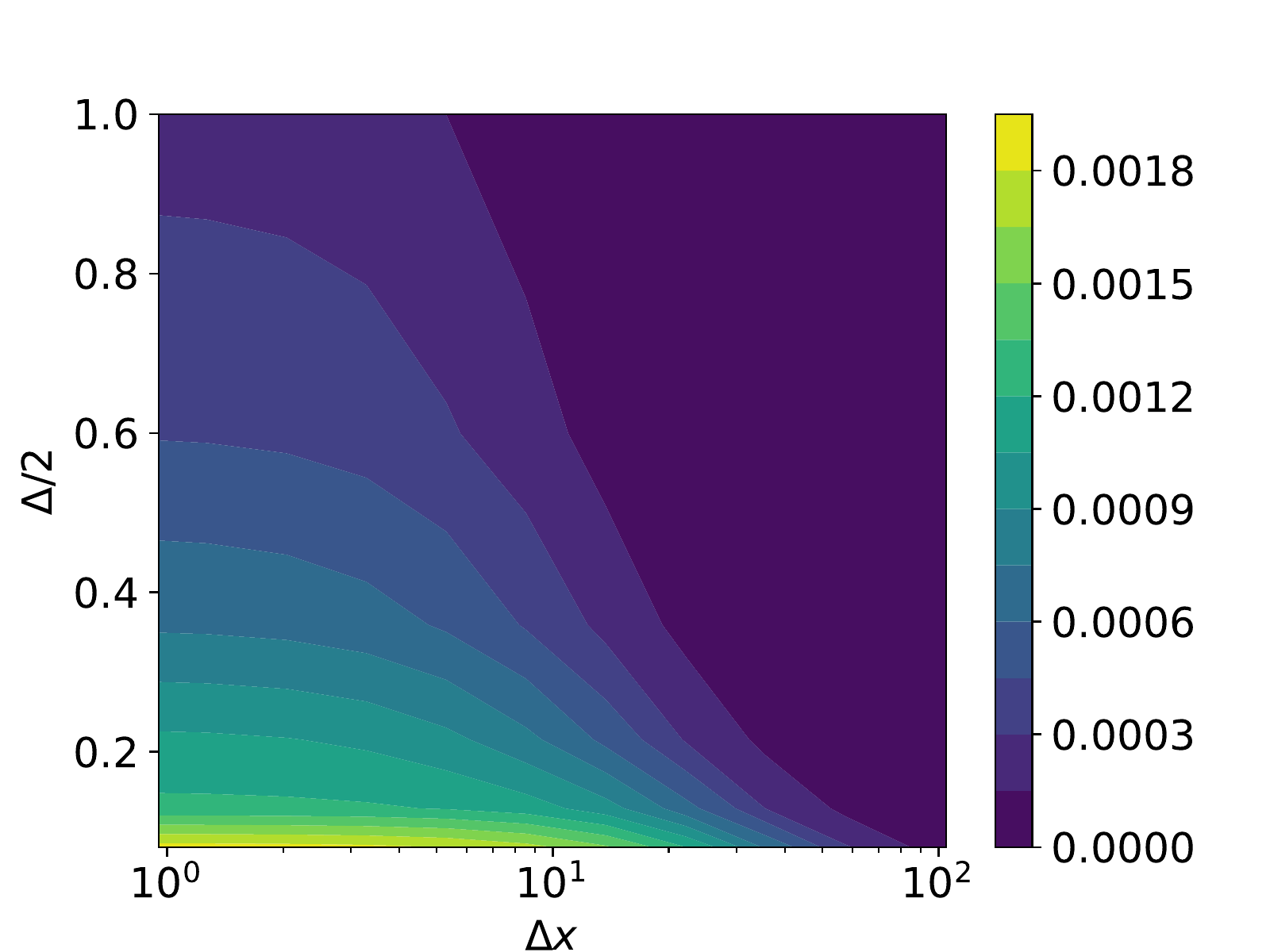}
    \includegraphics[scale = .36]{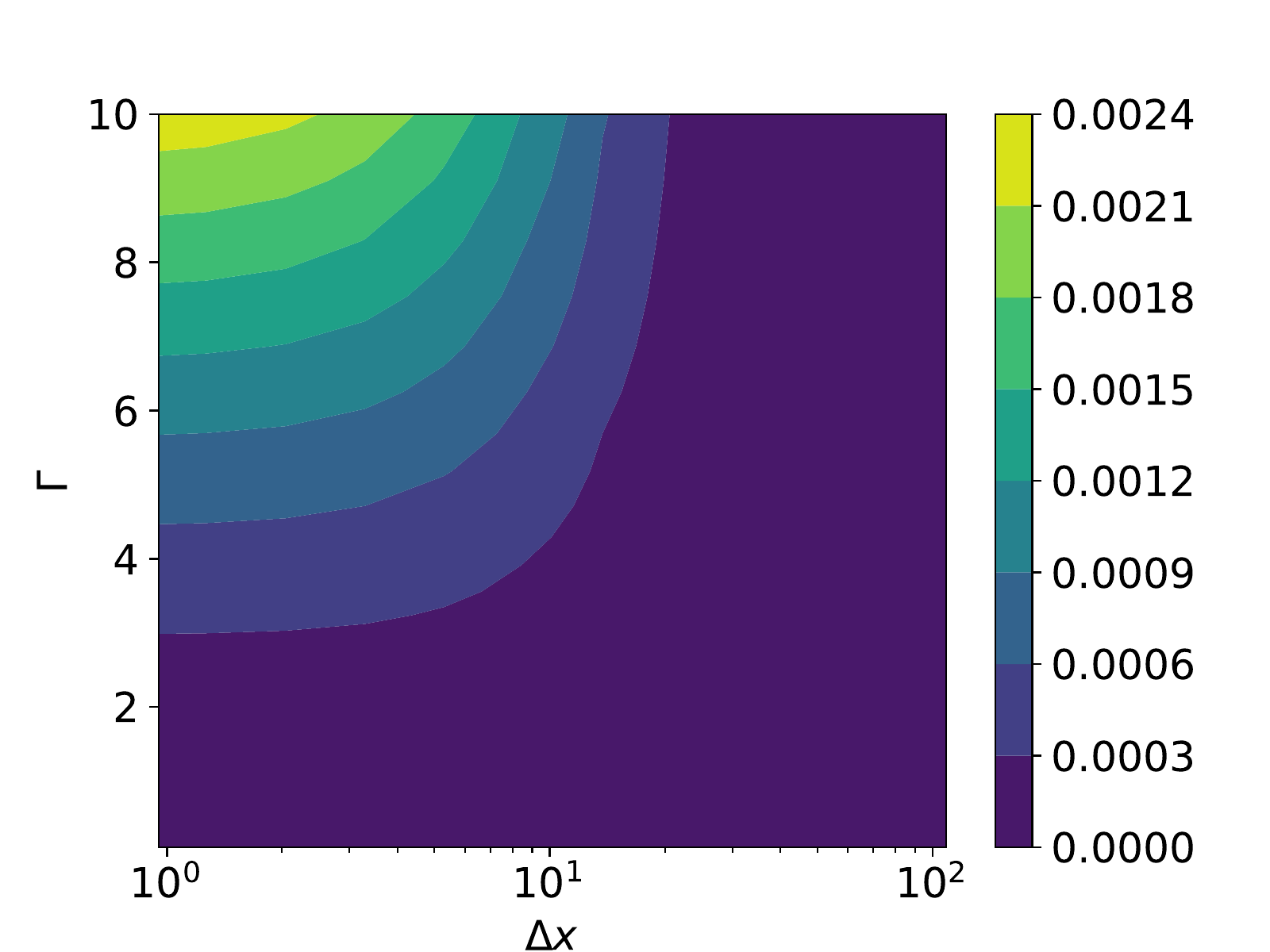}
    \caption{Modulus of the off-diagonal element of the steady state, $|\rho_{01}|$, from the numerical solution of \eqref{steady} in the case of a single qubit: {\em Left:} as a function of  $\Gamma$ and $\Delta$ for $\Delta x = 1$ . {\em Middle:} as a function of $\Delta x$ and $\Delta$ for $\Gamma = 5$.  {\em Right:} as a function of  $\Delta x$ and $\Gamma$ for $\Delta/2 = 0.6$. $\beta = 0.1$, $m = 0.1$,  $\lambda = L = \hbar = 1$ and $x_0 = -10$.}
    \label{fig:Coherences_2}
\end{figure*}

There are three key parameters that determine whether the qubit thermalizes or not: The Bohr energy $\Delta$, the bombardment rate $\Gamma$, and the dispersion of the particle's position $\Delta x$ that determine the magnitude of the off-diagonal terms of the density matrix $\rho_U$ in momentum representation, as shown by Eq.~\eqref{rhou}. The interplay among the three parameters is
not trivial, as shown in Fig.~\ref{fig:Coherences_2}, but is qualitatively captured by the following expression \cite{supp}:
\begin{equation}\label{estimation}
    |\rho_{01}|\sim  \frac{\Gamma}{\omega}    e^{-\beta m \Delta x^2\omega^2/2}
\end{equation}
where $\omega\equiv \Delta/\hbar$ is the Bohr frequency of the qubit. For a 1 GHz qubit, for example, one could observe coherence if effusion occurs at a rate of $10^9$ particles per second or greater, which is achievable \cite{Camposeo2001}   and compatible with the condition $\Gamma L\ll \sqrt{kT/m}$,  warranting that there is a single particle in the scattering region at any time (the presence of two or more particles could induce nonlinear effects that have been explored in the context of cavity QED \cite{Carmichael1999}).
The exponential factor in Eq.~\eqref{estimation} imposes a more involved condition on the spatial dispersion of the bombarding particles, but it does not seem very restrictive either: for the 1 GHz qubit, molecules with a mass of 1000 protons could generate coherences at room temperature if they are localized in an interval of the order of 50 nanometers or smaller.



To conclude, we have established the necessary conditions for thermalization in a wide class of quantum maps combined with a unitary evolution that lasts a random Poissonian time. We have determined when one could expect a deviation from thermalization due to quantum effects. Our results are useful to design repeated-interaction reservoirs that are thermodynamically consistent and could help to clarify current open problems. One example is whether the presence of non-commuting conserved observables hinders or boosts thermalization \cite{Halpern2022,Majidy2023,Kranzl2022}, a problem that has been partially addressed using generic collisional reservoirs \cite{Manzano2022}. In addition, they show that the system exhibits a high sensitivity to the kinetic characteristics of the bombarding particles in situations that can be reproduced in experimental setups, such as those used in cavity QED \cite{Haroche2006}, and could shed light on the problem of spatial decoherence of macromolecules \cite{Zeilinger2003,Arndt2004,Hornberger2003}.



JT-B and JMRP acknowledge financial support from 
the Spanish Government 
(Grant FLUID, PID2020-113455GB-I00) 
and from the Foundational Questions Institute Fund, a 
donor advised fund of Silicon Valley Community Foundation (Grant number FQXi-IAF19-01).
ME is funded by the Foundational Questions Institute Fund (Grant number FQXi-
IAF19-05). 
F. B. thanks Fondecyt project 1191441 and ANID – 
Millennium Science Initiative Program-NCN19-170.

\bibliography{collisional.bib}

\end{document}


\title{Thermalization and dephasing in  collisional reservoirs\\
Supplemental Material}

\author{Jorge Tabanera-Bravo}
\email{jorgetab@ucm.es}
\affiliation{Departamento de Estructura de la Materia, F\'isica T\'ermica y Electr\'onica and GISC, Universidad Complutense de Madrid, Pl. de las Ciencias 1. 28040 Madrid, Spain}

\author{Juan M. R. Parrondo}
\email{parrondo@ucm.es}
\affiliation{Departamento de Estructura de la Materia, F\'isica T\'ermica y Electr\'onica and GISC, Universidad Complutense de Madrid, Pl. de las Ciencias 1. 28040 Madrid, Spain}

\author{Massimiliano Esposito}
\email{massimiliano.esposito@uni.lu}
\affiliation{Complex Systems and Statistical Mechanics, Physics and Materials Science Research Unit, University of Luxembourg, L-1511 Luxembourg, G.D. Luxembourg}

\author{Felipe Barra}
\email{fbarra@dfi.uchile.cl }
\affiliation{Departamento de F\'isica, Facultad de Ciencias F\'isicas y Matem\'aticas, Universidad de Chile, 837.0415 Santiago, Chile}

\date{\today}

\maketitle

\setlength{\tabcolsep}{4pt}
\renewcommand{\arraystretch}{1.3}

\section{Transition amplitudes and simulations}

In this section, we summarize the methods used to solve the model, perform the simulations shown in Fig.~1, and obtain the steady state depicted in Figs.~2 and 3 of the main text.

We start by evaluating the scattering map $\mathbb{S}_{j'k'}^{jk}$ in Eq.~(17) of the main text. It contains the energy-dependent transition amplitudes $s_{j'j}^{(\alpha)}(E)$. In \cite{Tabanera2022} we provide an approximate expression for these quantities:
\begin{equation}
    s_{j'j}^{(+)}(E) \simeq
    \left\{
    \begin{split}
        &e^{-iL(k_{j'} + k_j)/2\bra{j'}e^{iL\mathbb{K}(E)}\ket{j}} \qquad &E \geq e_{\rm max},\\
        &\delta_{j'j} \qquad &E < e_{\rm max}.
    \end{split}\right.
\end{equation}
In this expression, the operator $\mathbb{K}(E) = \sqrt{2m(E-H_{\rm tot})}$ depends on the complete Hamiltonian $H_{\rm tot} = H_S + V$. The wave vectors are $k_j = \sqrt{2m(E-e_j)}$, and $e_{\rm max}$ is the maximum eigenvalue of $H_S$ and $H_{\rm tot}$. This approximation is valid for high incident kinetic energies and neglects the reflection amplitudes $s_{j'j}^{(-)}(E) \simeq 0$. With this aspproximation, we evaluate the scattering map $\mathbb{S}_{j'k'}^{jk}$ in Eq.(17) for the given $\rho_U(p,p')$ in the main text Eq.~(23).

The stochastic evolution in Fig.~1 of the main text is obtained by concatenating the scattering map with unitary evolution ${\mathbb U}^{\tau}\rho=e^{-iH_S\tau/\hbar}\rho e^{iH_S\tau/\hbar}$ with Poissonian distributed times $\tau$, and bombarding rate $\Gamma$. We initialize the system in a diagonal state with $\rho_{00} = 0$ and $\rho_{11} = 1$.

We obtain the results in Figs.~2 and 3 numerically solving Eq.~(9). In order to do this, we write Eq.~(9) in the eigenbasis of $H_S$,
\begin{equation}
    -\frac{i}{\hbar}\Delta_{jk}\rho_{jk} + \Gamma\sum_{j'k'}\left[\mathbb{S}_{j'k'}^{jk} - \delta_{j'j}\delta_{k'k}\right]\rho_{j'k'} = 0.
\end{equation}
For a single qubit density matrix, the last equation becomes a system of $2 \times 2$ linear equations that can be treated with standard techniques.

\section{Stationary density matrix}

Here, we obtain a rough approximation of the off-diagonal term $\rho_{01}$ of the stationary density matrix in the example discussed in the main text: a single qubit with energies $\pm\Delta/2$. According to Eqs.~(10) and (14), its modulus up to first order in $\Gamma$ reads 
\begin{equation}
    |\rho_{01}| \simeq \frac{\Gamma\hbar}{\Delta}|{\mathbb S}_{01}^{00}\rho^{(0)}_{00}+{\mathbb S}_{01}^{11}\rho^{(0)}_{11}|.
\end{equation}
Populations $\rho^{(0)}_{jj}$ are of order 1 and the two entries of the collisional map ${\mathbb S}_{01}^{00}$ and ${\mathbb S}_{01}^{11}$ are of the same order. Hence,
 the order of magnitude of $|\rho_{01}|$ is
 \begin{equation}
    |\rho_{01}| \sim \frac{\Gamma\hbar}{\Delta}|{\mathbb S}_{01}^{00}|.
\end{equation}
The term of the collisional map is given by Eq.~(17), which reads 
\begin{equation}\label{int1}
\mathbb{S}_{01}^{00} = \sum_{\alpha=\pm}\int_{ p_{\rm inf}}^\infty dp\, \rho_{U}(p,\pi(p))\sqrt{\frac{p}{\pi(p)}}\,s_{00}^{(\alpha )}\left(E_p+e_0\right)
\left[s_{10}^{(\alpha )}(E_{p}+e_{1})\right]^* 
\end{equation}
with $\pi(p)=\sqrt{p^2+2m\Delta}$. The density matrix of the particle in the momentum representation is given by Eq.~(23):
\begin{equation}
    |\rho_U(p,\pi(p))| =\mu_{\rm eff}\left(\frac{p+\pi(p)}{2}\right)\,\exp\left[
    -\frac{\Delta x^2\left[p-\pi(p)\right]^2}{2\hbar^2}\right].\label{rhousm1}
\end{equation}
The effusion distribution $\mu_{\rm eff}(p)$ is  peaked around its maximum at $p_{\rm max}=\sqrt{m/\beta}$. Then, the integral in \eqref{int1} is dominated  by  values of the momentum close to $ p_{\rm max}$. Furthermore, if $p_{\rm max}^2/m=1/\beta$ is much larger than $\Delta$, then
\begin{equation}
\pi(p_{\rm max})\simeq p_{\rm max}+\frac{m\Delta}{p_{\rm max}}.
\end{equation}
Hence
\begin{equation}\label{int2}
|\mathbb{S}_{01}^{00}| \sim \exp\left[
    -\frac{\Delta x^2  (m \Delta/p_{\rm max})^2}{2\hbar^2}\right]\times\left|\sum_{\alpha=\pm}s_{00}^{(\alpha )}\left(E_{p_{\rm max}}+e_0\right)
\left[s_{10}^{(\alpha )}(E_{p_{\rm max}}+e_{1})\right]^* \right|
\end{equation}


Assuming that the scattering amplitudes are not very sensitive to the momentum, we can replace the sum over $\alpha$ by a single dimensionless constant $s^2$. Then we get
\begin{equation}
   |\mathbb{S}_{01}^{00}|\sim  s^2\exp\left[
    -\frac{\Delta x^2  \left(m \Delta/(\sqrt{m/\beta}\right)^2}{2\hbar^2}\right]=s^2e^{-\beta m\Delta x^2\Delta^2/(2\hbar^2)}
\end{equation}
and
\begin{equation}\label{estimSM}
    |\rho_{01}| \sim  \frac{s^2\,\Gamma\hbar }{\Delta}    e^{-\beta m \Delta x^2\Delta^2/(2\hbar^2)}=
    \frac{s^2\,\Gamma}{\omega}    e^{-\beta m \Delta x^2\omega^2/2}
\end{equation}
where $\omega\equiv \Delta/\hbar$ is the Bohr frequency of the qubit.

\begin{figure}[h]
    \includegraphics[scale = .33]{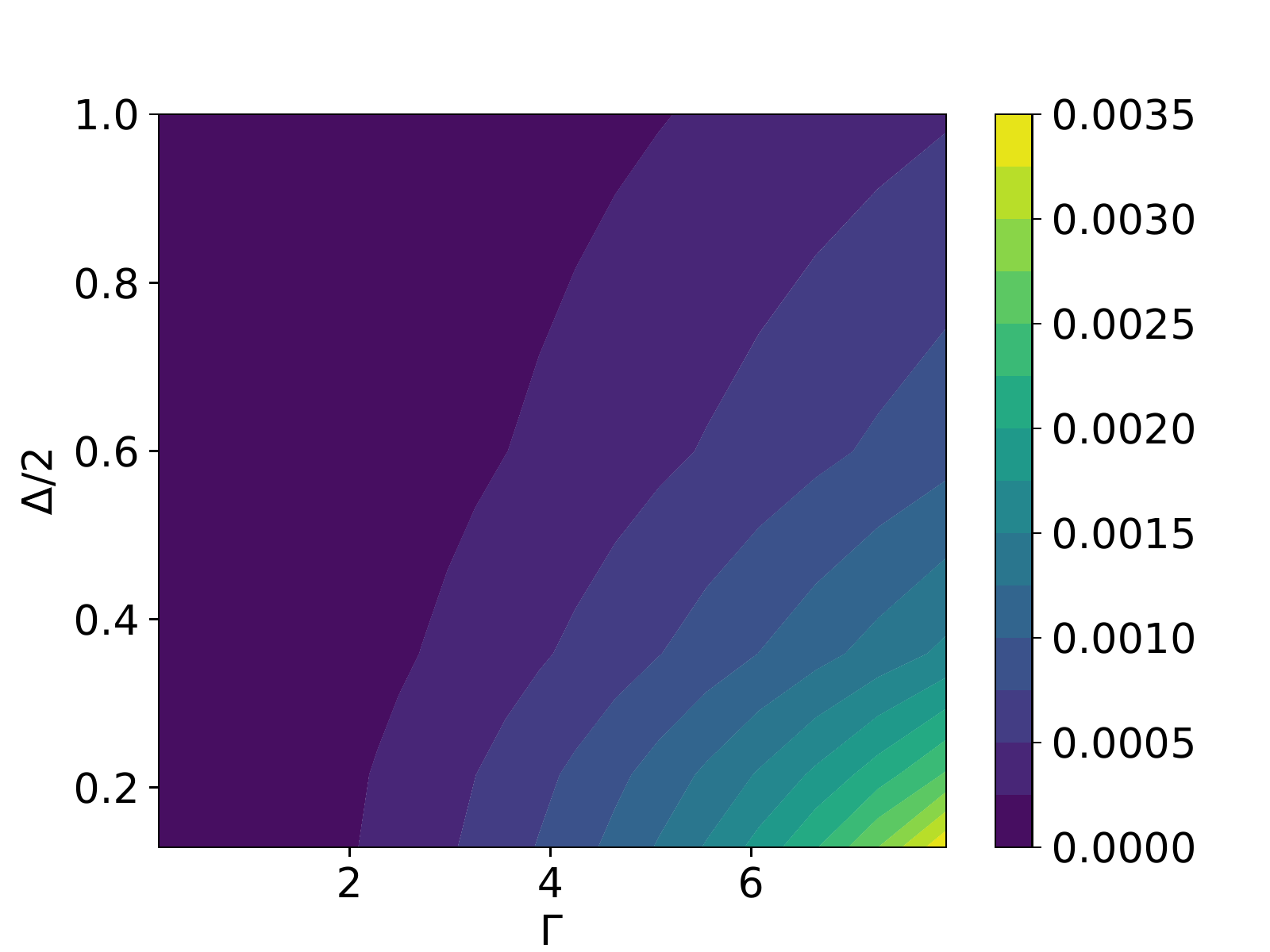}
    \includegraphics[scale = .33]{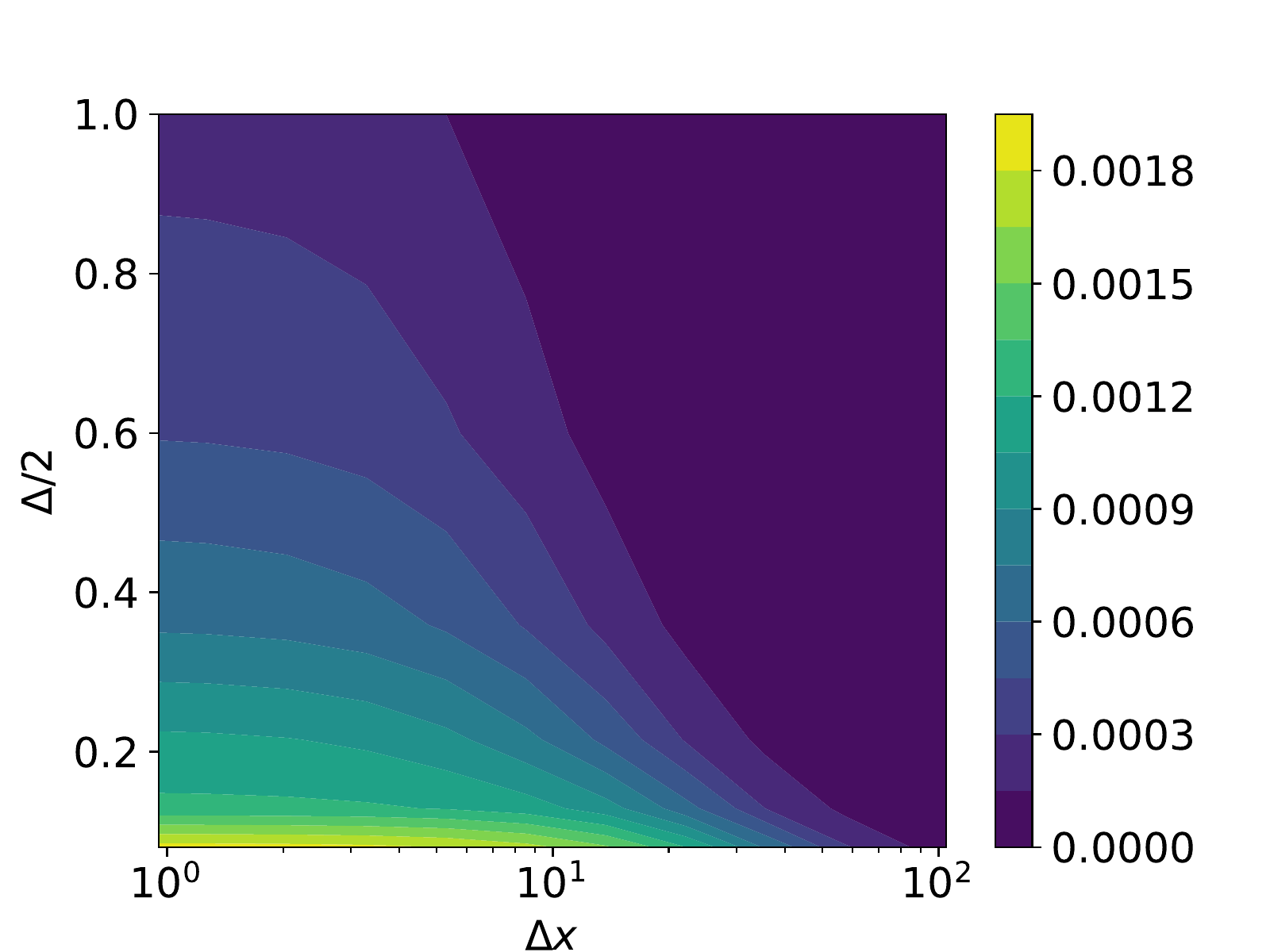}
    \includegraphics[scale = .33]{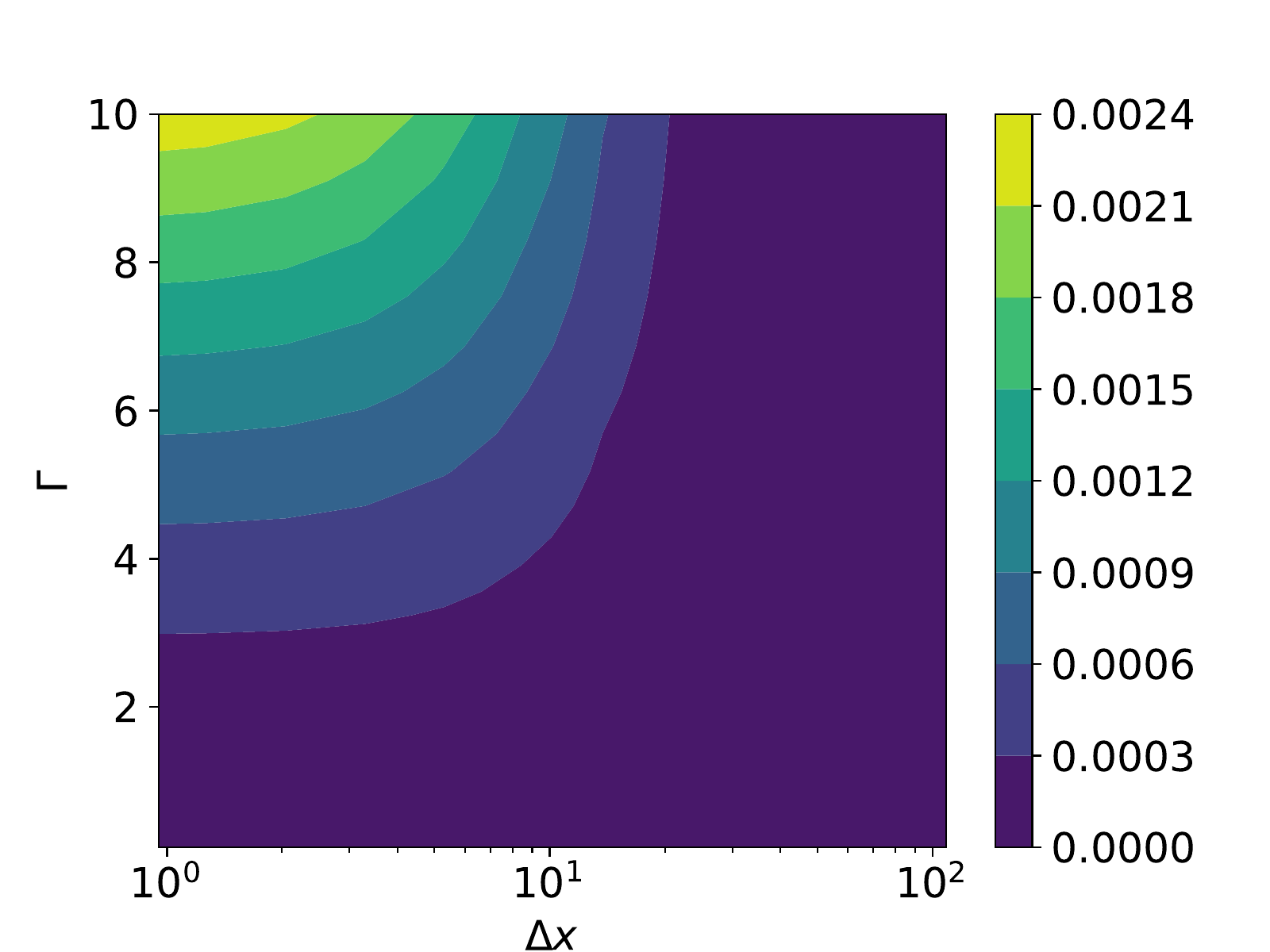}
    \caption{Off-diagonal element $|\rho_{01}|$ from the numerical solution of Eq.~(9) in the main text in the case of a single qubit (left) as a function of $\Delta$ and $\Gamma$ with $\Delta x = 1$ (middle)  as a function of $\Delta$ and $\Delta x$ with $\Gamma = 5$ and (right) as a function of $\Gamma$ and $\Delta x$ with $\Delta/2 = 0.6$.
    $\lambda = L = \hbar = 1$, $\beta = 0.1$, $m = 0.1$ and $x_0 = -10$.}
    \label{fig:Coherences_2SM}
\end{figure}

\begin{figure}[h]
    \includegraphics[scale = .38]{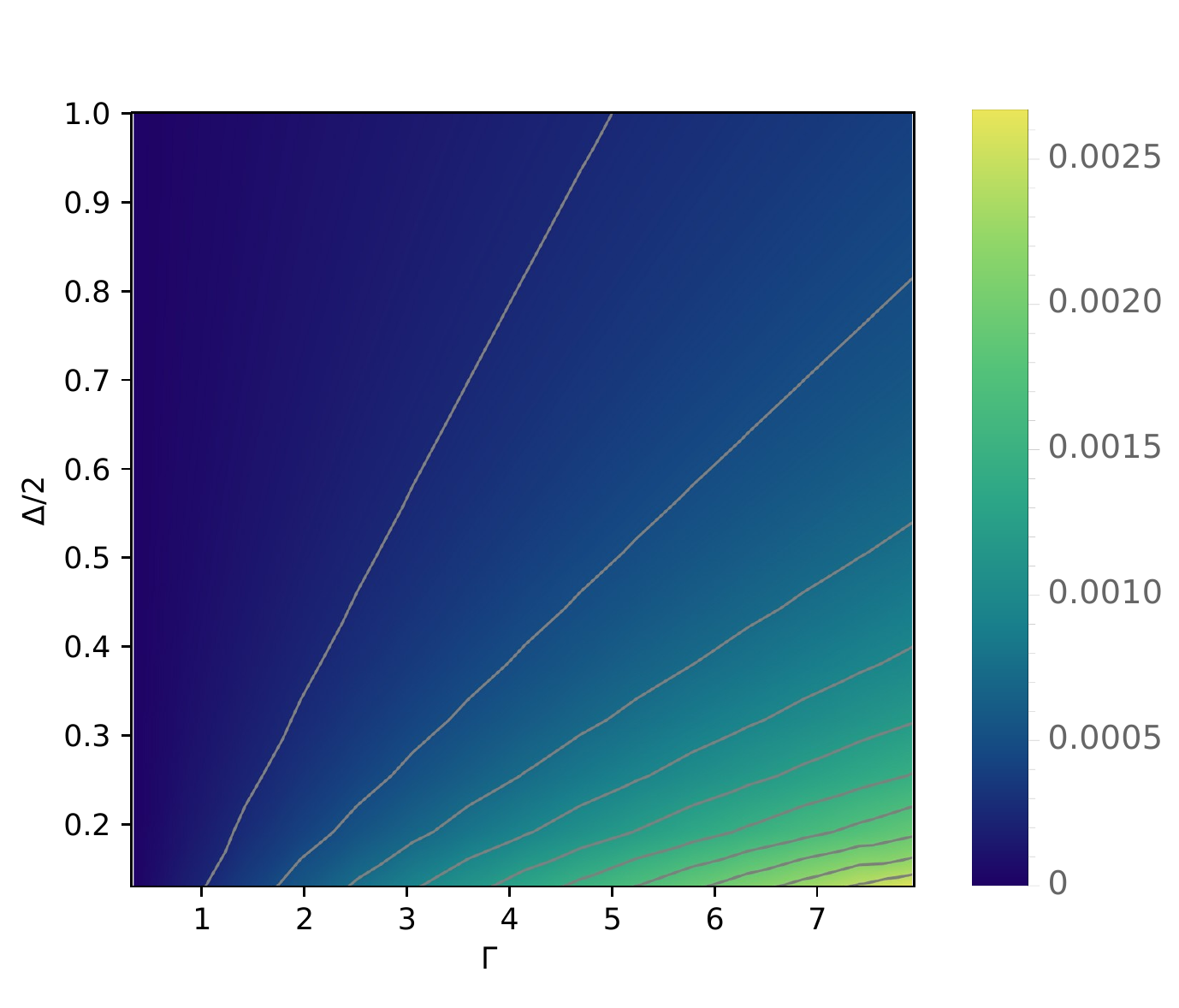}
    \includegraphics[scale = .38]{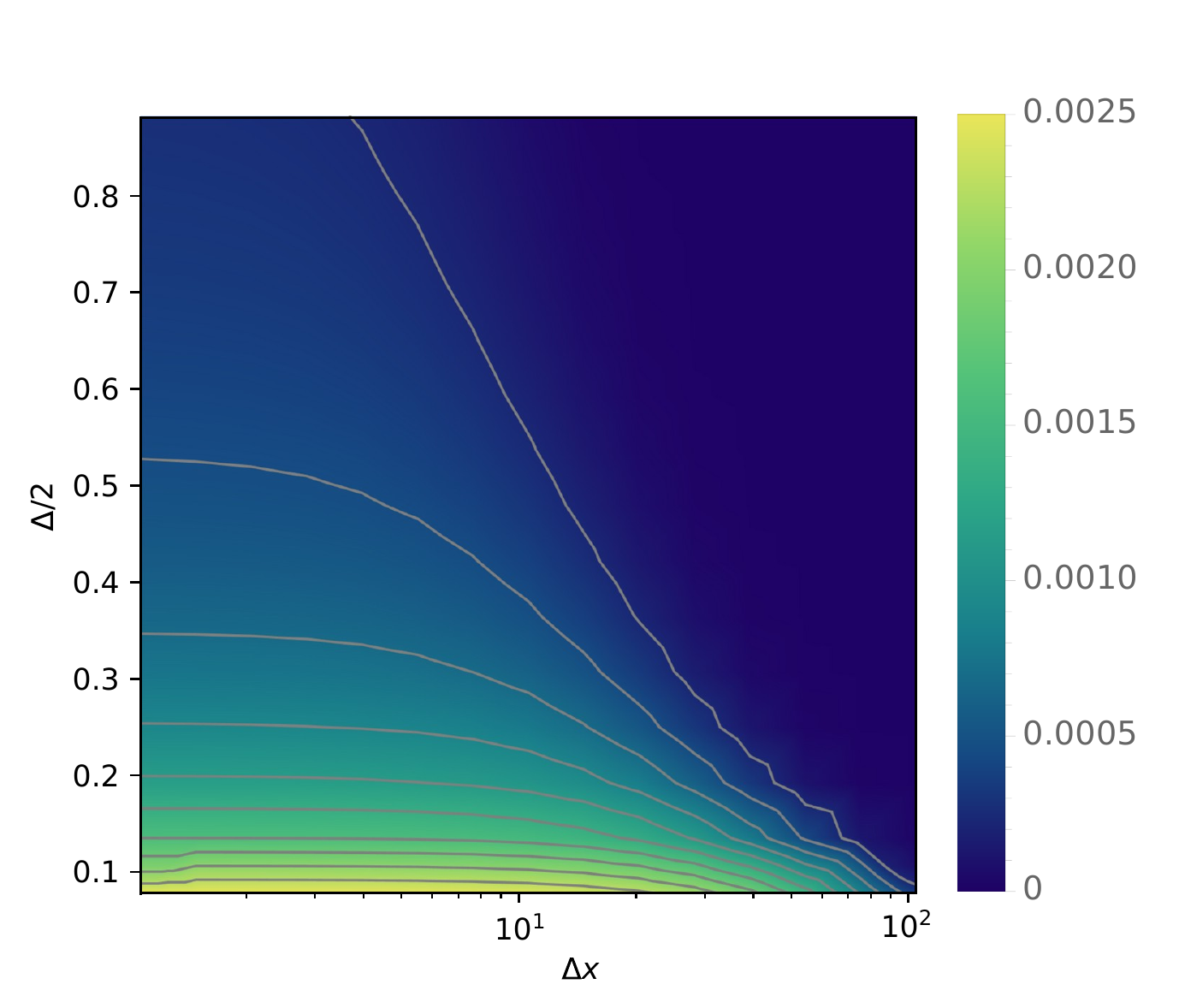}
    \includegraphics[scale = .38]{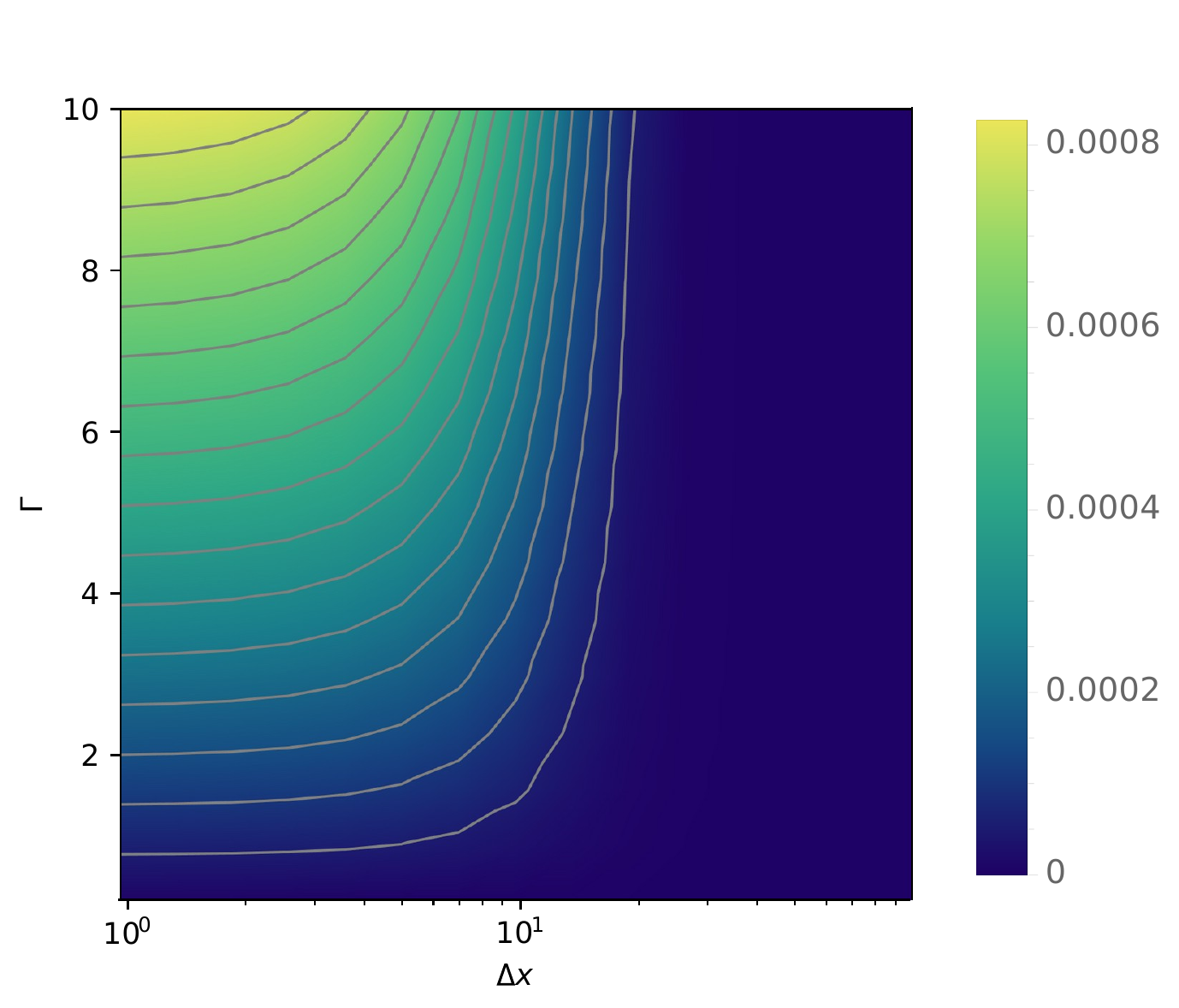}
    \caption{Off-diagonal element $|\rho_{01}|$ from the estimation \eqref{estimSM} and with $s=0.01$ and same parameters as in figure \ref{fig:Coherences_2SM}.
    }
    \label{fig:Coherences_3SM}
\end{figure}

In Figs.~\ref{fig:Coherences_2SM} and \ref{fig:Coherences_3SM}, we plot the value of $|\rho_{01}|$ in different situations obtained from the exact solution of Eq.~(9) in the main text, which gives the steady state for Poissonian bombardment at a rate $\Gamma$, and from the estimate given by \eqref{estimSM} with $s=0.01$. Although the estimation is very rough, it captures the qualitative behaviour of the off-diagonal term and provides a rule of thumb to determine whether the system thermalizes.

\vspace{1cm}

\bibliographystyle{apsrev4-1}

\bibliography{collisional.bib}